\documentclass[aps,pra,twocolumn,showpacs,superscriptaddress]{revtex4-1}
\usepackage{graphicx}
\usepackage{amsmath,amsfonts}
\usepackage[usenames]{color}

\newcommand{\tauQ}{\tau_{\rm Q}}

\def\be{\begin{equation}}
\def\ee{\end{equation}}
\def\bea{\begin{eqnarray}}
\def\eea{\end{eqnarray}}
\def\bi{\begin{itemize}}
\def\ei{\end{itemize}}

\begin{document}

\title{Thermal fluctuations and quantum phase transition in antiferromagnetic Bose-Einstein condensates}

\author{Emilia Witkowska}
\affiliation{Instytut Fizyki PAN, Aleja Lotnik\'ow 32/46, 02-668 Warsaw, Poland}

\author{Tomasz \'Swis\l{}ocki}
\affiliation{Instytut Fizyki PAN, Aleja Lotnik\'ow 32/46, 02-668 Warsaw, Poland}

\author{Micha\l{} Matuszewski}
\affiliation{Instytut Fizyki PAN, Aleja Lotnik\'ow 32/46, 02-668 Warsaw, Poland}

\begin{abstract}
We develop a method for investigating nonequilibrium dynamics of an ultracold system that is initially at thermal equilibrium. Our procedure is based on the classical fields approximation with appropriately prepared initial state. As an application of the method, we investigate the influence of thermal fluctuations on  the quantum phase transition from an antiferromagnetic to phase separated ground state in a spin-1 Bose-Einstein condensate of ultracold atoms. We find that 
at temperatures significantly lower than the critical condensation temperature $T_c$ the scaling law for the number of created spin defects remains intact.
\end{abstract}

\maketitle

\section{Introduction}
\label{sec_intro}

The ground state phase diagram of an antiferromagnetic Bose-Einstein condensate was studied experimentally in the regime where spatial and spin degrees of freedom are decoupled~\cite{GS_experiment},
demonstrating a phase transition from an antiferromagnetic phase (where only the $m_F=\pm1$ Zeeman components are populated) to a mixed broken axisymmetric phase 
(where all three Zeeman states can be populated). In this regime the system follows the predictions of the single spatial mode approximation (SMA). 
In other ranges of parameters, however, spatial separation of components can occur spontaneously,
and spin domains may appear in the ground state of the system~\cite{Matuszewski_AF, Matuszewski_PS}. 
In this case the ground state is either the antiferromagnetic phase or is phase separated into two domains,
and the two possibilities are divided by a critical point that is characterized by a critical magnetic field.

A system may become out of equilibrium when it is driven through the critical point 
due to the divergence of the relaxation time. 
If symmetry breaking occurs at the same time, this out-of-equilibrium process can produce various kinds of defects, 
depending on the dimensionality of the system and the form of the order parameter. 
The unified description of these phenomena is described by the Kibble-\.Zurek mechanism (KZM), 
which was studied in a number of physical systems, from the early Universe to ultracold atomic
gases~\cite{Kibble,other_KZ,Zurek,Helium_KZ,superconductors_KZ,BEC_KZ,Esslinger,Ulms}. 
Among these, Bose-Einstein condensates of ultracold atoms offer realistic models of highly controllable and tunable systems~\cite{Spinor_FerromagneticKZ}. 
Recent experiment with quasi-one-dimensional ultracold atoms confirmed the spontaneous creation of solitons via the Kibble-\.Zurek mechanism~\cite{Lamporesi}.

In recent papers~\cite{Nasz_PRL, Nasz_PRB}, we demonstrated that the quantum phase transition from an 
antiferromagnetic to phase separated ground state in a spin-1 Bose-Einstein condensate of ultracold atoms 
exhibits scaling laws characteristic for systems displaying universal behavior. 
Phase separation leads to the formation of spin domains, with the number of domain walls depending on the quench time. 
Interestingly, the Kibble-\.Zurek scaling law was confirmed only for the dynamics close to the critical point. 
Further evolution led to postselection of domains, which gave rise to a second scaling law 
with a different exponent. The postselection was attributed to the conservation of an additional quantity, 
namely the condensate magnetization.

In this paper, we develop a method for investigating nonequilibrium dynamics of an ultracold system 
that is initially at thermal equilibrium. 
We apply this method to investigate the effect of nonzero temperature on the dynamics of 
the phase transition and the resulting scaling laws.
We consider this problem within the framework of the classical theory of a complex field 
with exact equations of motion being the Gross-Pitaevskii equations~\cite{CFM}.
The initial condition for the transition includes both the condensate and thermal atoms that 
introduce thermal fluctuations in the system.
We sample the initial thermal equilibrium within the Bogoliubov approximation at a given 
temperature and for fixed total number of atoms and magnetization.
Modes orthogonal to the condensate are thermally populated 
according to the Bogoliubov transformation.
In our work, the field has to be interpreted not simply as the condensate 
wavefunction, but rather as the total matter field. 
We present both the results of single realizations of the field, which experimentally 
correspond to single experimental runs, and results of averaging over different initial states.
We find that while the dynamics of the system can be altered by the thermal fluctuations, 
at relatively low temperatures the scaling law for the number of domains in the final state is intact.

\section{The model}
\label{sec_model}

In the following, we consider a dilute, weakly interacting spin-1 Bose-Einstein condensate 
placed in a homogeneous magnetic field pointing along the $z$ axis.
For the sake of completeness, we recall the details of the model used previously in~\cite{Nasz_PRL,Nasz_PRB} to describe the 
dynamics of a condensate at zero temperature within the truncated Wigner approximation. In the following section, we
will generalize this approach to the case of a condensate at nonzero initial temperature.

We start with the Hamiltonian $H = H_0 + H_{\rm A}$, where the symmetric (spin-independent) part is 
\begin{equation} \label{En}
H_0 = \sum_{j=-,0,+} \int d x \, \psi_j^\dagger \left(-\frac{\hbar^2}{2m}\nabla^{2} + \frac{c_0}{2} \rho 
+ V(x)\right) \psi_j. 
\end{equation}
Here the subscripts $j=-,0,+$ denote sublevels with magnetic quantum numbers along the magnetic field axis $m_f=-1,0,+1$,
$m$ is the atomic mass, $\rho=\sum \rho_j = \sum \psi_j^\dagger \psi_j$ is the total atom density, 
$V(x)$ is the external potential. 
Here we restricted the model to one dimension, with the other degrees of freedom confined
by a strong transverse potential with frequency $\omega_\perp$. The spin-dependent part can be written as
\begin{equation} \label{EA}
H_{\rm A} = \int d x \, \left[ \sum_j E_j \rho_j + \frac{c_2}{2} :{\bf F}^2:\right]\,,
\end{equation}
where $E_j$ are the Zeeman energy levels, the spin density is 
${\bf F}=(\psi^{\dagger}F_x\psi,\psi^{\dagger}F_y\psi,\psi^{\dagger}F_z\psi)$,
where $F_{x,y,z}$ are the spin-1 matrices and $\psi =(\psi_+,\psi_0,\psi_-)$.
The spin-independent and spin-dependent interaction coefficients are given by 
$c_0=2 \hbar \omega_\perp (2 a_2 + a_0)/3>0$ and $c_2= 2 \hbar \omega_\perp (a_2 - a_0)/3>0$, 
where $a_S$ is the s-wave scattering length for colliding atoms with total spin $S$.
In the following analytic calculations we assume the incompressible regime where
\be 
c_0~\gg~|c_2|~,
\ee
which is a good approximation eg.~in the case of $^{87}$Rb or $^{23}$Na spin-1 condensate.

The total number of atoms $N = \int \rho d x$ and magnetization $M = \int \left(\rho_+ - \rho_-\right) d x$ are conserved quantities. 
In reality, there are processes that can change both $N$ and $M$,
but they are relatively weak both in spin-1 $^{23}$Na and $^{87}$Rb condensates~\cite{Chang_NP_2005,deHaas} 
and can be neglected on the time scales considered below.

The linear part of the Zeeman shifts $E_j$ induces a homogeneous rotation of the spin vector around the direction of the magnetic field.
Since the Hamiltonian is invariant with respect to such spin rotations, 
we consider only the effects of the quadratic Zeeman shift~\cite{Matuszewski_AF,Matuszewski_PS}.
For sufficiently weak magnetic field we can approximate it by a positive energy shift of the $m_f=\pm 1$ sublevels 
$\delta=(E_+ + E_- - 2E_0)/2 \approx B^2 A$, 
where $B$ is the magnetic field strength and $A=(g_I + g_J)^2 \mu_B^2/16 E_{\rm HFS}$,
$g_I$ and $g_J$ are the gyromagnetic ratios of electron and nucleus, $\mu_B$ is the Bohr magneton, 
$E_{\rm HFS}$ is the hyperfine energy splitting at zero magnetic field \cite{Matuszewski_AF,Matuszewski_PS}.
Finally, the spin-dependent Hamiltonian (\ref{EA}) becomes
\begin{equation} 
H_{\rm A} = \int d x \, \left[ AB^2(\rho_+ + \rho_-) +\frac{c_2}{2} :{\bf F}^2: \right].
\end{equation}

\begin{figure}
\includegraphics[width=8.5cm]{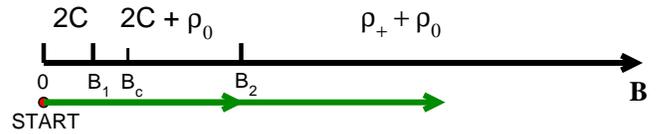}
\caption{Ground state phase diagram of an antiferromagnetic condensate for magnetization $M=N/2$.
The green arrow indicates the direction of quench into a phase separated state
considered in Sec.~\ref{sec_kzm}.
}
\label{scenario}
\end{figure}

Except for the special cases $M=0,\pm N$, the $V(x)=0$ ground state phase diagram, shown in Fig.~\ref{scenario}, 
contains three phases divided by two critical points at
\begin{equation}
B_1=B_0 \frac{M}{\sqrt{2} N},~~
B_2=B_0 \frac{1}{\sqrt{2}},
\end{equation} 
where $B_0=\sqrt{c_2 \rho/A}$ and $\rho$ is the total density. The ground state can be 
$(i)$ antiferromagnetic (2C) with $\psi=(\psi_+,0,\psi_-)$ for $B < B_1$,
$(ii)$ phase separated into two domains of the 2C and $\psi=(0,\psi_0,0)$ type ($\rho_0$) for $B\in(B_1, B_2)$, or
$(iii)$ phase separated into two domains of the $\rho_0$ and $\psi=(\psi_+,0,0)$ type ($\rho_+$) for $B>B_2$~\cite{Matuszewski_PS}.
What is more, the antiferromagnetic 2C state remains dynamically stable, i.e., it remains a local energy minimum up to a critical 
field $B_c>B_1$. Consequently, the system driven adiabatically from the 2C phase, across the phase boundary $B_1$, and into the 
separated phase remains in the initial 2C state up to $B_c>B_1$ when the 2C state becomes dynamically unstable towards
the phase separation.

\section{Modeling thermal fluctuations by the classical fields}
\label{sec_thermal}

Classical fields method is used for description of thermal effects of a single condensate \cite{CFM}. 
Here we extend the applicability of the method to the system of spin-1 Bose-Einstein 
condensates with antiferromagnetic interactions.

In the method the condensed and noncondensed parts of the $j$-th component are described by a complex function $\psi_j$.
The description takes into account modes $k$ with energies lower than the temperature:
\bea
\frac{\hbar ^2 k_{max}^2}{2 m} n_{\rm cut} \simeq k_B T \, .
\eea
The relation between maximal momentum $k_{max}$ and the temperature is not strictly defined and $n_{\rm cut}$
has no unique value~\cite{my}. In general, results of the classical field are cut-off dependent. 
Here, we assume that $n_{cut}=1/3$.

In our simulations, we consider a lattice model for a classical field $\psi_j(x)$ with lattice spacing ${\rm dx}$. We enclose the atomic field in the 
ring-shaped quasi-1D geometry with periodic boundary conditions at $\pm L/2$ and $V(x)=0$. The
total number of atoms is a constant of motion
\bea
{\rm dx} \sum_{x,j} |\psi_j(x)|^2 = N \, ,
\eea
as well as the total magnetization 
\bea
{\rm dx} \sum_x \left( |\psi_+(x)|^2 -  |\psi_{-}(x)|^2 \right) = M \, .
\eea
The numerical evolution of the field is governed by discretized counterpart of the Hamiltonian (\ref{En}), (\ref{EA}).

The initial antiferromagnetic (2C) state at thermal equilibrium is prepared by employing Bogoliubov transformation of state 2C:
$\psi_j = \tilde{\psi}_j + \delta \psi_j$~\cite{Nasz_PRB}, with the constrain $\tilde{\psi}_0=0$. Linearization of the Gross-Pitaevskii equation in small fluctuations 
$\delta\psi_j(t,x)$ around uniform background $\tilde{\psi}_j$ decouple fluctuations $\delta \psi_\pm$ from $\delta \psi_0$.

Fluctuations $\delta \psi_\pm$ are composed of phonon $(p)$ and magnon $(m)$ branches
\bea\label{BogolPM}
&&
\left(
\begin{array}{c}
\delta\psi_+ \\
\delta\psi_-
\end{array}
\right)= \\
&&
\left(
\begin{array}{c}
\sqrt{\rho_+} \\
\sqrt{\rho_-}
\end{array}
\right)
\sum_{k\ne 0}^{k_{max}}
~\left(
b_k^{(p)} u_k^{(p)} e^{ikx}+
b_k^{(p)*} v_k^{(p)*} e^{-ikx}
\right)
~+~\nonumber\\
&&
\left(
\begin{array}{c}
\sqrt{\rho_-} \\
-\sqrt{\rho_+}
\end{array}
\right)
\sum_{k\ne 0}^{k_{max}}
\left(
b_k^{(m)} u_k^{(m)} e^{ikx}+
b_k^{(m)*} v_k^{(m)*} e^{-ikx}
\right),\nonumber
\eea
with quasiparticle energies
\bea 
\epsilon_k^{(p)}&=&c_2\rho\sqrt{\xi_{\rm s}^2k^2\left[2(c_0/c_2)+\xi_{\rm s}^2k^2\right]},~~ \nonumber\\
\epsilon_k^{(m)}&=&c_2\rho\sqrt{\xi_{\rm s}^2k^2(8 n_+ n_-+\xi_s^2k^2)}.
\label{eq:en_pm}
\eea
Normalized modes satisfy
\bea 
u_k^{(p)}\pm v_k^{(p)}&=&\left(\frac{\xi_{\rm s}^2k^2}{2(c_0/c_2)+\xi_{\rm s}^2k^2}\right)^{\pm1/4}~,\nonumber\\
u_k^{(m)}\pm v_k^{(m)}&=&\left(\frac{\xi_{\rm s}^2k^2}{8n_+ n_- +\xi_{\rm s}^2k^2}\right)^{\pm1/4}~,
\eea
where $n_\pm=\rho_\pm/\rho$. Here we use the spin healing length $\xi_{\rm s}=\hbar/\sqrt{2mc_2\rho}$.

The small quadrupole mode fluctuations $\delta \psi_0$ ~\cite{Ho_PRL_1998} are given by
\be \label{Bogol0}
\delta\psi_0~=~
\sum_{k=0}^{k_{max}}
\left(
b_k^{(0)} u_k^{(0)} e^{ikx} + b_k^{(0)*} v_k^{(0)*} e^{-ikx}
\right)
\ee
with their gapped spectrum for $b<b_c$
\be 
\epsilon_k^{(0)}~=~c_2\rho\sqrt{[\xi_{\rm s}^2k^2+(1-b^2)]^2-(1-b_c^2)^2}
\label{epsilonk0}
\ee
and the normalized eigenmodes 
\be 
u_k^{(0)}\pm v_k^{(0)}~=~
\left(
\frac{(b_c^2-b^2)+\xi_{\rm s}^2k^2}{2(1-b_c^2)+(b_c^2-b^2)+\xi_{\rm s}^2k^2}
\right)^{\pm1/4} .
\ee
Here we use a rescaled dimensionless magnetic field
\be 
b~=~\frac{B}{B_0}~.
\ee 

To generate the stochastic initial values of the classical field we proceed as follows. 
(i) For each realization, we generate the fluctuations $\delta \psi_j$ at temperature $T$ by 
introducing thermal population of the Bogoliubov modes. 
In practice we generate complex numbers $b_k^{({\rm x})}$ for $k\ne0$ 
of $j=\pm$ components and for all $k$ of $j=0$ component according to the probability distribution
\begin{equation}
P(b_k^{({\rm x})})= \frac{1}{\pi} \frac{\epsilon_k^{({\rm x})}}{k_B T} e^{-|b_k^{({\rm x})}|^2/k_B T} \, .
\end{equation}
Here $({\rm x})$ denotes $(m),\, (p)$ or $(0)$.
For a given realization, we build up fluctuations $\delta \psi_j$ according to Bogoliubov transformations (\ref{BogolPM}) and (\ref{Bogol0}).
(ii) Then, we create the classical field with the constraint that the total atom number and magnetization
are fixed. The form of the field is $\psi_j = \frac{a_j}{\sqrt{L}} + \delta \psi_j$ with
\bea
a_+ &=& \left( \frac{N-M -N_\perp}{2} \right)^{1/2} , \\
a_- &=& \left( \frac{N+M -N_\perp}{2} \right)^{1/2} , 
\eea
where $N_\perp=\sum_{k, j} |\delta \psi_j|^2$ is the number of non condensed atoms.

The ensemble  of classical fields created in this way is the initial state for the
time-dependent Gross-Pitaevskii (GP) equations
\bea \label{Wigner}
i\hbar\frac{\partial \psi_0}{\partial t}  &=& \left(-\frac{\hbar^2 \nabla^2}{2m} + c_0\rho\right)\psi_0 + \nonumber\\
      &&+c_2\left[(\rho_++\rho_-)\psi_0+2\psi_0^*\psi_+\psi_-\right],\nonumber\\
i\hbar\frac{\partial \psi_+}{\partial t}   &=& \left(-\frac{\hbar^2 \nabla^2}{2m} +c_0\rho+AB^2\right)\psi_+ + \nonumber\\
      &&+c_2\left[(\rho_+-\rho_-)\psi_++\rho_0\psi_++\psi_-^*\psi_0^2\right],\\
i\hbar\frac{\partial \psi_-}{\partial t}   &=& \left(-\frac{\hbar^2 \nabla^2}{2m} +c_0\rho+AB^2\right)\psi_- + \nonumber\\
      &&+c_2\left[(\rho_--\rho_+)\psi_-+\rho_0\psi_-+\psi_+^*\psi_0^2\right].\nonumber
\eea
One of the advantages of the method is the ability to calculate various correlation functions 
in a straightforward way, taking averages over many realizations of the stochastic fields.

Application of the method is justified for a very low temperature only since Bogoliubov transformation is 
the approximate solution. Moreover, very early evolution of the GP equations 
can display some transient effects due to the fact that 
the Bogoliubov transformation used in the sampling does not produce an exactly stationary distribution.
We have checked if such transients occur for parameters used in our simulations and found them to be marginal for
the quantities we are interested in.

\section{Validity of the Bogoliubov transformation}
\label{sec_val}

\begin{figure*}[]
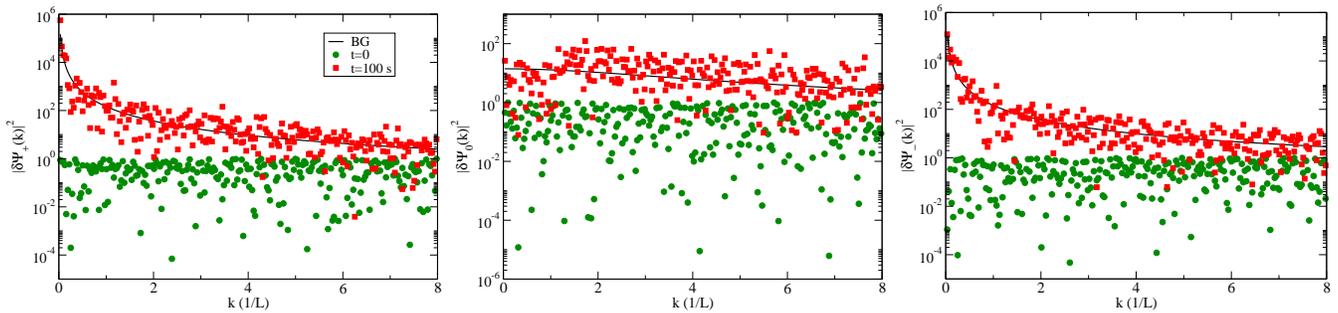

\begin{center}
\includegraphics[width=5.8cm]{dpsiP.eps}
\includegraphics[width=5.8cm]{dpsi0.eps}
\includegraphics[width=5.8cm]{dpsiM.eps}
\caption{Momentum distribution of non-condensed fields $|\delta \psi_j(k)|^2$ for $m_F=1$ (left panel), $m_F=0$ (central), $m_F=-1$ (right). Green: initial distribution; Red: after $100$s of equilibration with GPEs, Black: Bogoliubov transformation Eqs.(\ref{BogolPM}) and (\ref{Bogol0}). Here $N=10^7$, $M=N/2$, $b=0$, with length $L=200\,\mu$m and $\omega_\perp=2\pi \times 1000\,$Hz, other parameters as for sodium.}
\label{occupation_ran}
\end{center}
\end{figure*}

\begin{figure*}[]
\includegraphics[width=5.8cm]{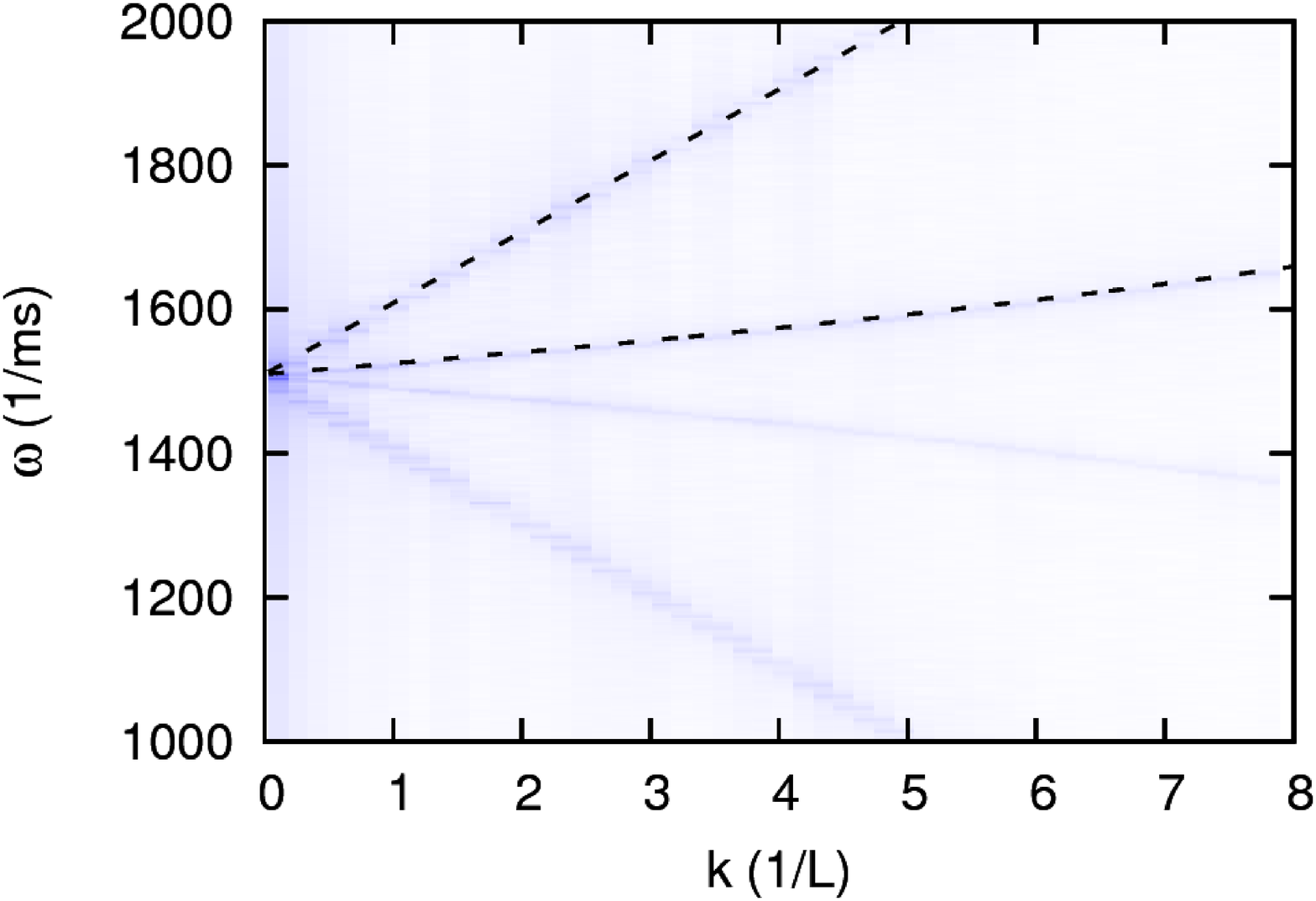}
\includegraphics[width=5.8cm]{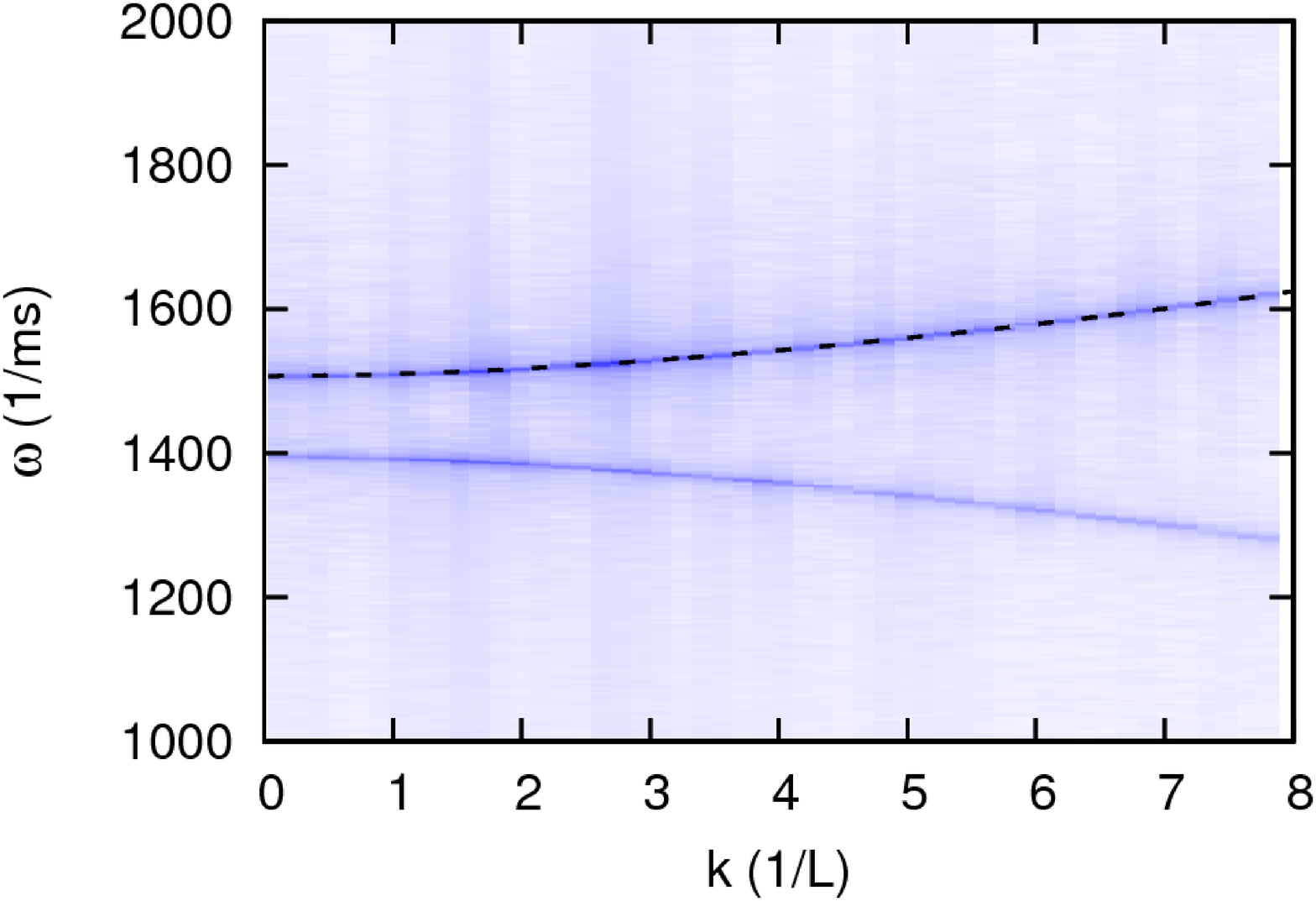}
\includegraphics[width=5.8cm]{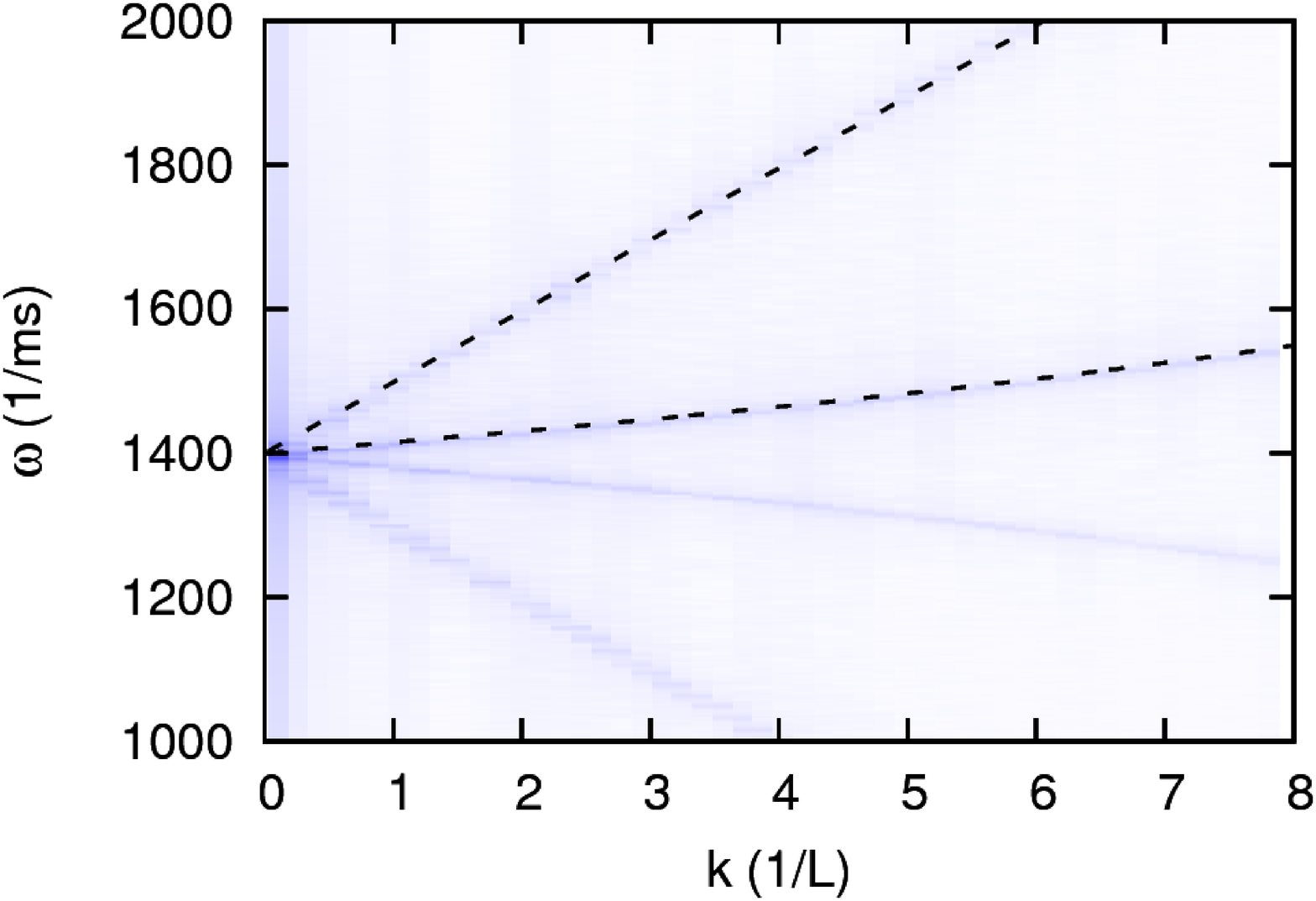}
\caption{Spectral density $|\psi_j(k,\omega)|^2$, for $m_F=1$ (left panel), $m_F=0$ (central), $m_F=-1$ (right) after $100$s of evolution with GPEs. 
Black dashed lines: quasiparticle energies (\ref{eq:en_pm}) and (\ref{epsilonk0}) given by the Bogoliubov theory. 
Parameters of the simulation are the same as in fig.~\ref{occupation_ran}.}
\label{spectrum_ran}
\end{figure*}

The dynamics with Gross-Pitaevskii equations leads the two component antiferromagnetic initial state to equilibrium state for long enough times. The equilibration given by this coupled set of nonlinear equations is surprising because we known they may lead to the coherent off-equilibrium spin-mixing dynamics. Indeed, it was shown by using mean-field theory and adapting the single-spatial-mode approximation, that the condensate dynamics is well described by a nonrigid pendulum and displays a variety of periodic oscillations~\cite{SpinMixing}. Fortunately, the period of oscillations depends on the initial fraction of atoms in the $m_F=0$ component and is almost infinite in our case. Therefore, the time scale associated with spin-multimode dynamics is much shorter and allows for equilibration of the system.
The relaxation observed in the system is quite intriguing in particular in the context of prethermalization phenomena~\cite{prethermalization}. 

In the low temperature limit, the Bogoliubov theory well describes the equilibrium state of the system. The comparison of the result of equilibration with GP equations to the Bogoliubov theory can be treated as an independent test of the last one. To this end, we checked the validity of the Bogoliubov transformation, (\ref{BogolPM}) and (\ref{Bogol0}), as well as the quasiparticle energies, (\ref{eq:en_pm}) and (\ref{epsilonk0}).
We start numerical calculations with randomly chosen initial fluctuations with flat distribution in momentum space, see green points in fig.~\ref{occupation_ran}, and superimposed constraints of given norm $N$ and magnetization $M$. Then we let the initial state to evolve with Gross-Pitaevskii equations for a transient time ($100$s for results presented in Figs.~\ref{occupation_ran} and~\ref{spectrum_ran}). Next, we compare the achieved distribution of non-condensed field in momentum space $\delta \psi_j(k\ne 0)$ to the predictions of the Bogoliubov theory. The comparison shown in Fig.~\ref{occupation_ran} is satisfactory. Analysis in the frequency domain allows for a test of the validity of the elementary excitation picture, see density plots in fig.~\ref{spectrum_ran}. One can easily recognize the magnon and phonon branches for $m_F=\pm1$ and the quadrupole branch for $m_F=0$. Once again, the comparison with the Bogoliubov theory (marked by black dashed lines) is satisfactory and shows that the quasiparticle energies (\ref{eq:en_pm}) and (\ref{epsilonk0}) are recovered by equilibration applied to the initial state.

\section{The Kibble-\.Zurek mechanism}
\label{sec_kzm}

In~\cite{Nasz_PRL,Nasz_PRB} we investigated the phase transition from the antiferromagnetic
to the phase separated state in a spin-1 Bose-Einstein condensate.
This continuous phase transition is driven by the change of the magnitude of the applied magnetic field.
Due to the spatial symmetry breaking in the phase separated state, the transition is accompanied by
the creation of multiple defects in the form of spin domain walls, with the number of domains dependent on the quench time.
The concept of the Kibble-\.Zurek mechanism (KZM) relies on the fact that the system does not follow the ground state exactly in the 
vicinity of the critical point due to the divergence of the relaxation time. 
The dynamics of the system cease to be adiabatic at 
$t\simeq-\hat t$
(here we choose $t=0$ in the first critical point),
when the relaxation time becomes comparable to the inverse quench rate
\begin{equation} \label{tau_freeze}
\hat{\tau}_{\rm rel} \approx |\hat{\varepsilon} / \hat{\dot{\varepsilon}}|,
\end{equation}
where 
$\varepsilon(t)=B - B_c\sim t/\tauQ$
is the distance of 
the system from the critical point.
At this moment, the fluctuations approximately freeze, until the relaxation time
becomes short enough again. After crossing the critical point, distant parts of the system choose 
to break the 
symmetry in different ways, which
leads to the appearance of multiple defects in the form of 
domain walls between domains of 
2C and $\rho_0$ phases. The average number of
domains is related to the correlation length 
$\hat\xi$ 
at the freeze out time 
$\hat{t} \sim \tauQ^{z\nu / (1+ z\nu)}$~\cite{Zurek,Nonequilibrium}
\begin{equation} \label{KZ_scaling}
N_{\rm d} = L / \hat{\xi} \sim \tauQ^{-\nu/(1+z\nu)},
\end{equation}
where $z$ and $\nu$ are the critical exponents determined by the scaling of the relaxation time 
$\tau_{\rm rel} \sim |\varepsilon|^{-z\nu}$ and excitation spectrum $\omega \sim |k|^z$, with $z=1$ in the superfluid.

Interestingly,
the Kibble-\.Zurek scaling law gives correct predictions only for the dynamics close to the critical point. Further on,
the post-selection of domains was observed, which gave rise to a second scaling law with a different exponent.
The post-selection was attributed to the conservation of an additional quantity, namely the condensate magnetization.

The analytical and numerical calculations of~\cite{Nasz_PRL,Nasz_PRB} were carried out within 
the zero-temperature limit of the truncated Wigner approximation.
In this section we use the method of Sec.~\ref{sec_thermal} to estimate 
the influence of nonzero temperature
on the dynamics of the phase transition and the resulting scaling laws.
We find that while the dynamics of the system can be
altered by the thermal fluctuations, at relatively low temperatures
the scaling law for the number of domains in the final state is intact.

\begin{figure}[tbp]
\includegraphics[width=8.5cm]{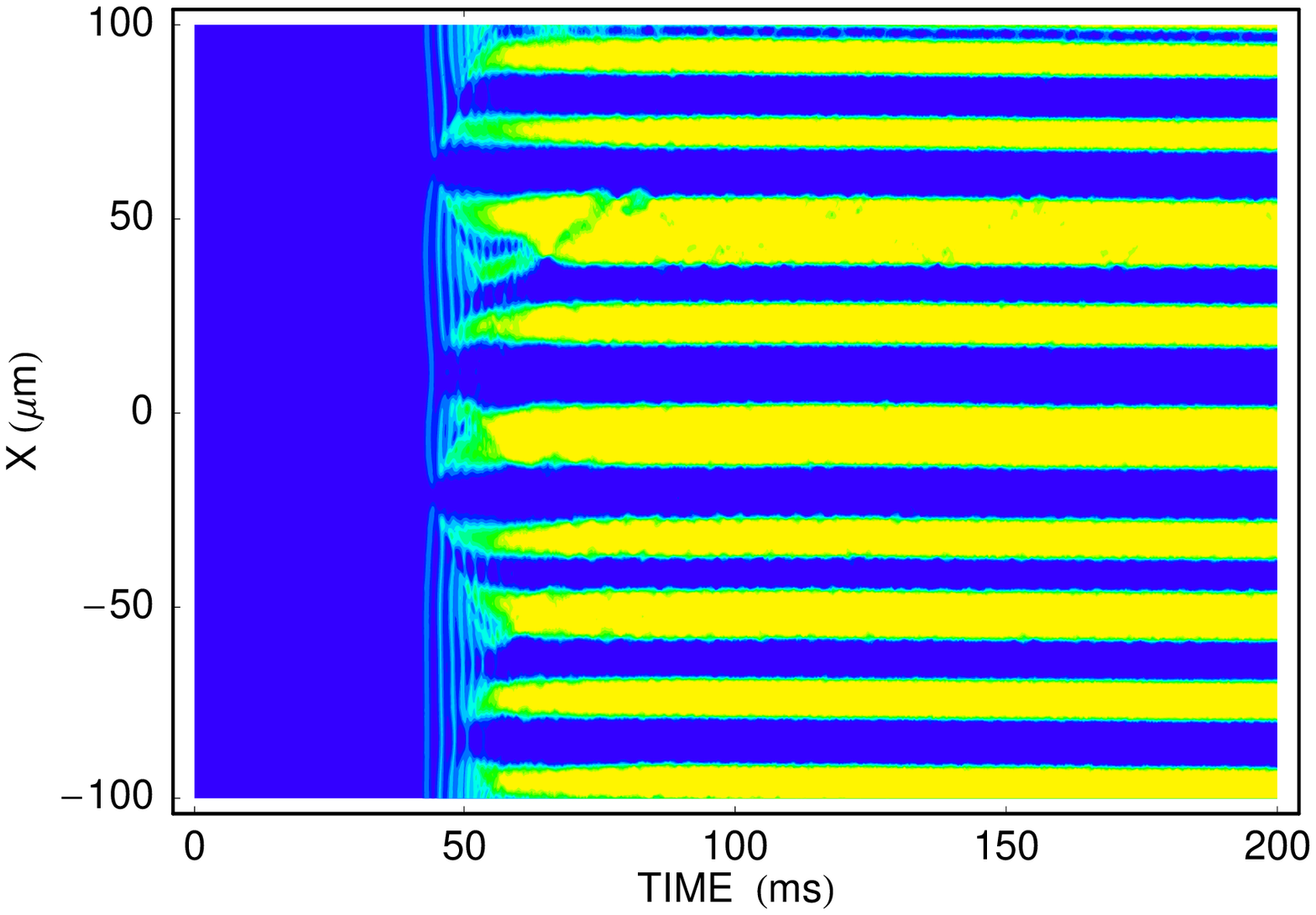}
\includegraphics[width=8.5cm]{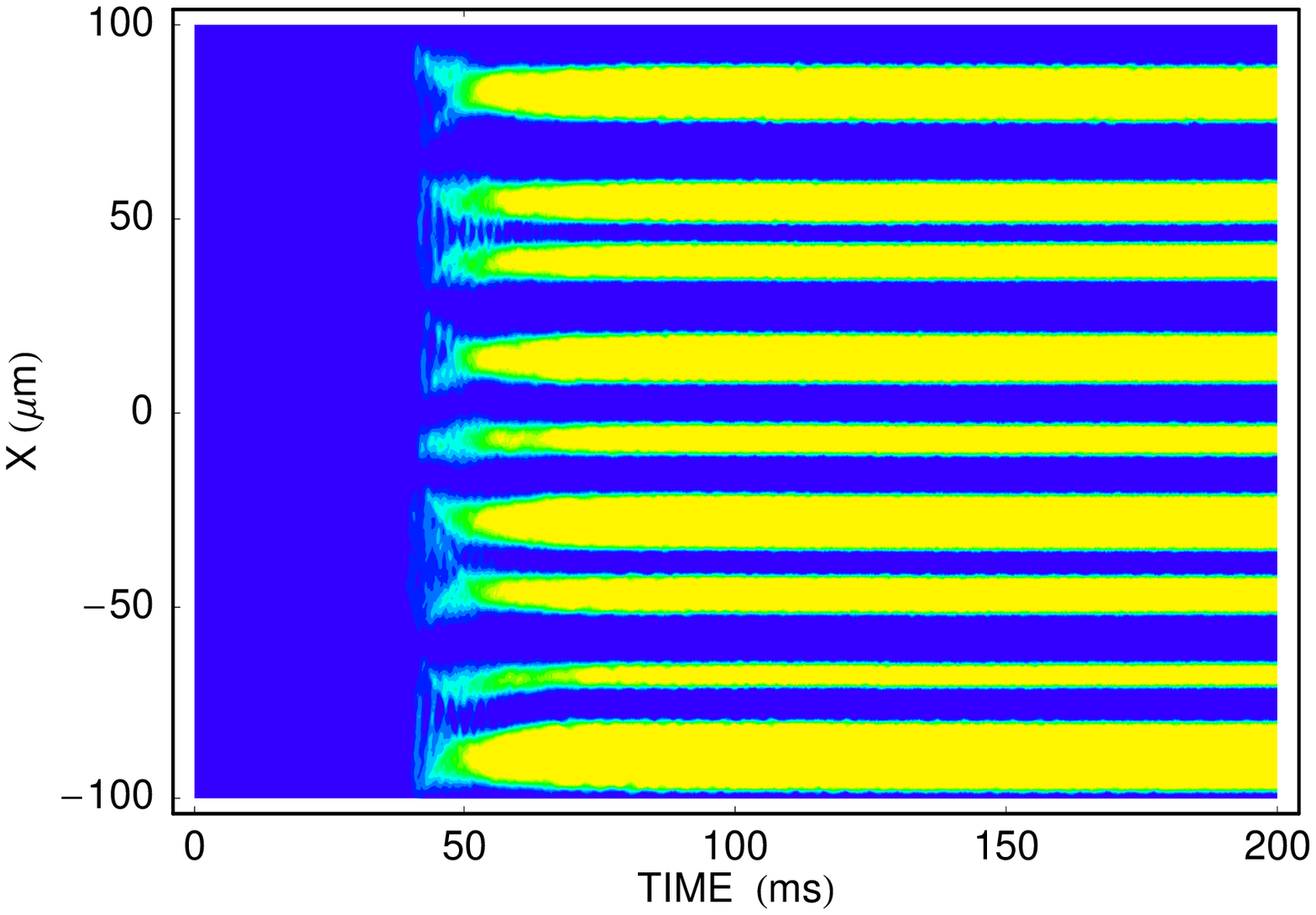}
\caption{
Formation of spin domains through a modified Kibble-Zurek mechanism 
in a ring-shaped 1D geometry with ring length $L=200\,\mu$m and $\omega_\perp=2\pi \times 1000\,$Hz,
for $N=10^{6}$ atoms. The evolution of the density of the $m_f=0$ component $|\psi_0|^2$ in a single realization of the experiment 
is shown. 
The top and bottom panels correspond to the zero temperature and finite temperature results. The quench time is taken as $\tauQ=100\,$ms. 
Total condensed fraction at $t=0$ is $N_{k=0}/N=0.89$.
}
\label{fig_snapshots}
\end{figure}

\begin{figure}[tbp]
\includegraphics[angle=-90,width=\columnwidth]{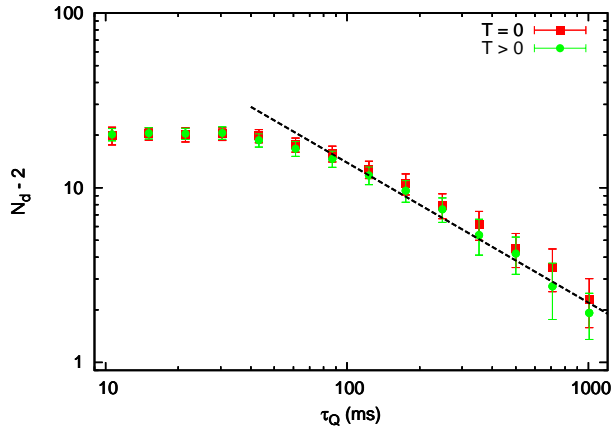}
\caption{Comparison of the averaged number of spin domains $N_{\rm d}$ after
the quench as a function of the quench time for $N=10^6$ atoms.
The points are results of numerical simulations
averaged over 100 runs.
The square and circle points correspond to zero temperature and finite temperature results.
The scale is logarithmic on both axes and $N_{\rm d}$ is decreased by two
to account for the ground state phase separation into two
domains. The dashed line is the fit to the power
law with scaling exponent $n_d=-0.73\pm0.04$. The grid size is $2^8$ and the cut-off is equal to 1/3.
}
\label{fig_scalings}
\end{figure}

We now describe in detail the scenario of the experiment. The antiferromagnetic spin-1 condensate is trapped in a 
ring-shaped quasi-1D trap with strong transverse confinement and the circumference length $L$.
The magnetic field is initially switched off, and the atoms are prepared in the homogeneous 
antiferromagnetic (2C) ground state with magnetization 
set to $M=N/2$. To investigate KZM we increase $B$ linearly as
\begin{equation}
B(t)~=~B_0~\frac{t}{\tauQ},
\label{tauQ}
\end{equation}
to drive the system through the two phase transitions into a phase separated state. Then, at $t>\tauQ$
the magnetic field is kept constant at the level $B=B_0$.
As described in Sec.~\ref{sec_model}, the ground state of the system becomes separated
into the 2C and the $\rho_0$ phase at $B=B_1$, but the initial 2C state remains metastable
until the point $B=B_c$. 
At this point, the system is expected to undergo phase transition accompanied by spatial symmetry breaking. 
As described above, due to the finite quench time the phase transition has a nonequilibrium
character, and multiple spin domains can develop in the system, instead of two as the form of the ground state would suggest.
Further, at the second critical point $B=B_2$, 
there is no symmetry breaking accompanying the phase transition and the spin-domain landscape remains intact.
The mean-field critical exponents of the symmetry-breaking phase transition 
are $z=1$ and $\nu=1/2$, which according to the formula
(\ref{KZ_scaling}) gives the scaling law for the number of domains as $N_d\sim \tau_{\rm Q}^{1/3}$. However,
as shown in~\cite{Nasz_PRL,Nasz_PRB}, this prediction is correct only for the number domain seeds formed close to the critical point.
When the domains become fully formed, their number is decreased in the post-selection process, which is due
to the existence of an upper limit for the number of domains in a system with conserved magnetization $M$.

Here, we investigate the influence of the finite temperature on the process described above. 
In Fig.~\ref{fig_snapshots} we show the evolution of the density of atoms in the initially unoccupied $m_f=0$ state
during the process of domain formation. The top figure shows a single realization of the zero-temperature truncated Wigner
simulation, which can be interpreted as a result of a single experiment. For comparison, analogous 
result in the case of finite temperature, obtained using the method described in Sec.~\ref{sec_thermal}, is shown in
the figure below. While this particular example corresponds to a relatively short quench time, when the post-selection process is not very
effective, it is visible that some of the initial fluctuations merge to form a single domain instead of two.
The effect of the finite temperature can be seen as the picture of the dynamics being more ``fuzzy'' in the figure below,
but the number of created domains and their properties seem to be unaffected due to the relatively low temperature. 

This conclusion is further confirmed in Fig.~\ref{fig_scalings}, where we show the results of systematic averaging
of the number of created domains over many realizations of the initial distribution. The two sets of points
showing the result for $T=0$ and finite temperature overlap except for the very long times, where some decrease of the number of
domains in the $T>0$ case can be observed. Close inspection of the dynamics leads us to account this slight decrease 
(on average by less than one domain) to the lower number of domains created at the symmetry phase transition
when the initial state already contains thermal fluctuations.
In particular, the finite temperature does not affect significantly the $N_d\sim \tau_{\rm Q}^{2/3}$ scaling law predicted
to result from the domain post-selection process. This turns out to be the case for any temperature
investigated by our method  based on the Bogoliubov transformation. 
We note that the scaling law may be affected by thermal effects at temperatures higher than the ones 
achievable using the current method. This will be the topic of a future study.

\section{Conclusions}
\label{sec_summary}

In summary, we developed a method for investigating nonequilibrium dynamics of an ultracold system that is initially at thermal equilibrium. Our procedure is based on the classical fields approximation with appropriately prepared initial state.  
We described in details how to model thermal fluctuations using the Bogoliubov transformation, and
demonstrated its validity by performing dynamical equilibration 
of an antiferromagnetic initial state with given total atom number and magnetization. 
We studied the effect of the non-zero temperature on the scaling law for the number of domains created via the Kibble-\.Zurek mechanism in antiferromagnetic spinor condensates. The effect of the finite temperature is visible on the evolution of the gas density. We find the density to be more "fuzzy" but the number of created domains 
and their properties rather unaffected by the temperature. This result in a relatively low temperature shows the strength of the conservation law, namely the conservation of magnetization in our system that enforce the scaling law.
In addition to the main result, we observed relaxation of the antiferromagnetic initial state to 
a thermal-like equilibrium. This intriguing feature provides an interesting direction for future work.

\acknowledgments

This work was supported by the Polish Ministry of Science and Education grant IP 2011 034571, the National Science Center grant DEC-2011/03/D/ST2/01938, and by the Foundation for Polish Science through the \textquotedblleft Homing Plus\textquotedblright\ program.

\clearpage

\end{document}